\documentclass[11pt,a4paper]{article}
\usepackage{jheppub}
\pdfoutput=1
\usepackage{graphicx}
\usepackage{bm}% bold math
\usepackage{amssymb}
\usepackage{caption}
\usepackage{subcaption}
\usepackage{paralist}
\usepackage{multirow}
\usepackage[textsize=scriptsize,backgroundcolor=red!40,linecolor=red]{todonotes}

\newcommand{\paperkeyw}{Prompt neutrino, QCD, IceCube}

\usepackage[all]{hypcap}

\newcommand{\be}{\begin{eqnarray}}
\newcommand{\beq}{\begin{equation}}
\newcommand{\eeq}{\end{equation}}
\newcommand{\ee}{\end{eqnarray}}

\newcommand{\bmp}{\noindent\begin{minipage}{16cm}}
\newcommand{\emp}{\end{minipage}\vskip 7mm} % 7mm untightened

\newcommand{\gev}{\rm GeV}

\newcommand{\nue}{\ensuremath{\nu_{e}}}
\newcommand{\numu}{\ensuremath{\nu_{\mu}}}

\newcommand{\kmcb}{\ensuremath{\text{km}^{3}}}

\newcommand{\dzero}{\ensuremath{D^{0}}}
\newcommand{\dpm}{\ensuremath{D^{\pm}}}

\newcommand{\ccb}{\ensuremath{c\bar{c}}}

\newcommand{\zmom}{\ensuremath{Z\text{-moments}}}

\newcommand{\eg}{\textit{e.g.}}
\newcommand{\ie}{\textit{i.e.}}

\newcommand{\diff}{\ensuremath{\mathrm{d}}}
\newcommand{\ddiff}{\ensuremath{\mathrm{d}^2}}

\title{Perturbative charm production and the prompt atmospheric neutrino flux in light of RHIC and LHC}
 
\preprint{NORDITA-2015-9}

\author[a]{Atri Bhattacharya,}
\author[b]{Rikard Enberg,}
\author[c]{Mary Hall Reno,}
\author[a,d]{Ina Sarcevic}
\author[e,f]{and Anna Stasto}
\affiliation[a]{Department of Physics, University of Arizona,\\1118 E. Fourth Street, Tucson, AZ, 85721 U.S.A.}
\affiliation[b]{Department of Physics and Astronomy, Uppsala University,\\Box 516, Uppsala, 751 20 Sweden}
\affiliation[c]{Department of Physics and Astronomy, University of Iowa,\\30 North Dubuque Street, Iowa City, IA, 52242 U.S.A.}
\affiliation[d]{Department of Astronomy and Steward Observatory, University of Arizona,\\933 North Cherry Avenue, Tucson, AZ, 85721 U.S.A.}
\affiliation[e]{Department of Physics, Pennsylvania State University,\\University Park, PA 16802, U.S.A}
\affiliation[f]{Institute of Nuclear Physics, Polish Academy of Science,\\ul.\ Radzikowskiego 152, Krak\'ow, Poland}
\emailAdd{atrib@email.arizona.edu}
\emailAdd{rikard.enberg@physics.uu.se}
\emailAdd{mary-hall-reno@uiowa.edu}
\emailAdd{ina@physics.arizona.edu}
\emailAdd{astasto@phys.psu.edu}

\abstract{We re-evaluate the prompt atmospheric neutrino flux, using the measured charm cross sections at RHIC and the Large Hadron Collider to constrain perturbative QCD parameters such as the factorization and renormalization scales, as well as modern parton distribution functions and recent estimates of the cosmic-ray spectra.
We find that our result for the prompt neutrino flux is lower
than previous perturbative QCD estimates and, consequently, alters the
signal-to-background statistics of the recent IceCube
measurements at high energies.
}


\arxivnumber{1502.01076}

\begin{document}
\maketitle

\section{Introduction}

Decays of mesons produced when cosmic ray primaries at high energies interact with
nuclei in the atmosphere lead to the production of a large flux of neutrinos.
The resulting neutrino flux at energies from 10 GeV--1 TeV has been observed by several
pioneering experiments \cite{Ahlen:1995av, Daum:1995bf, Fukuda:1998ub, Alekseev:1998ib,
Hatakeyama:1998ea, Sanchez:2003rb, Adamson:2005qc, Adamson:2012gt} over the last two decades.
The dominant contribution to the neutrino flux at these energies comes from
the decays of charged pions
$ \pi^{\pm} \to \mu^{\pm} \numu \to e^{\pm} \nue \numu \numu$, and from leptonic
and semi-leptonic decays of kaons \cite{Gaisser:2009gn}.
Of late, the construction of first the AMANDA detector \cite{Andres:1999hm}
and then, its upgrade to the first $ \gtrsim 1\,\kmcb $ volume neutrino telescope,
the IceCube (IC) Neutrino Observatory, has made the detection of neutrinos at much higher energies
$ E_{\nu} \gtrsim 100 $ TeV possible (see \cite{Halzen:2006mq} for a review). Moreover, there are plans for the KM3NeT detector in the Mediterranean \cite{Katz:2014aoa}, and for an IceCube-Gen2 upgrade~\cite{Aartsen:2014njl}, which will reach even higher neutrino energies.
While AMANDA had already seen atmospheric neutrinos in the high TeV's \cite{Abbasi:2010qv},
IC has recently revealed its first observation of high energy neutrinos with
energies going up to 2.1 PeV \cite{IceCube, Aartsen:2014gkd}.
In total, IC observes 37 events in the energy range 10 TeV--2.1 PeV over 988 days,
with atmospheric neutrinos and muons expected to be responsible for about
10--11 of those,
their fluxes being especially significant for energies up to 200 TeV.
A precise understanding of the prompt neutrino background at these energies is
crucial in determining the IC signal statistics. Because present values for the flux used
by IC depend on theoretical results\footnote{IceCube has recently published an upper limit \cite{Aartsen:2014muf} on the prompt flux at energies above 1 TeV at less than 1.52 times the flux predicted in ref.\ \protect\cite{ers}.} \cite{Volkova:1980sw,Gondolo:1995fq,prs,Martin:2003us,ers}, it is timely to
revisit the theoretical predictions in light of updated information on the various QCD
inputs as well as recent high-energy experimental results such as those obtained at the Large Hadron Collider (LHC).

Prompt neutrinos, produced in the atmosphere by decays of charmed hadrons
(\eg, $ \dzero, \dpm \to \nu_{e,\mu} X $) that come from cosmic-ray interactions with
atmospheric nuclei ($ pN \to c\bar{c}X $), become an important component of the
atmospheric neutrino flux at these high energies \cite{Volkova:1980sw,Gondolo:1995fq,prs,Martin:2003us,ers}.
For the IC study of neutrinos from astrophysical sources,
atmospheric neutrinos are a background and therefore, need to be carefully
estimated.
The neutrino flux from pions and kaons, hereafter called the conventional neutrino flux, has
been previously calibrated against multi-GeV atmospheric neutrino observations. This conventional flux
is well understood, but the prompt neutrino flux, being sub-dominant
at lower energies, is less so.

The evaluation of the prompt flux requires as a first step the computation of
the charm production cross-section in the process $ pN \to c\bar{c}X $, followed
by the {hadronization} of charm particles.
The computation of charm production in hadronic collisions introduces some of the biggest
uncertainties that translate into the evaluation of the prompt flux.
These arise due to uncertainties in the knowledge of
\begin{inparaenum}[\itshape a\upshape)]
	\item the charm mass ($ m_c $),
	\item the factorization ($ M_F $) and renormalization ($ \mu_R $) scales, and
	\item the choice of parton distribution functions (PDF's).
\end{inparaenum}
In addition, the final step in computing the prompt lepton flux involves
folding the charmed hadron production cross-section with the flux of
incoming cosmic-rays, and thus incurs large uncertainties from the
limited understanding of the extremely high-energy cosmic ray composition.

Two approaches to the calculation of the prompt lepton flux are via perturbative QCD
(pQCD) in the parton model (as in, \eg, \cite{Gondolo:1995fq,prs,Martin:2003us}) and with the dipole model
(see, \eg, \cite{ers}).
Recently, the production of charm mesons in the atmosphere has also
been studied using the event generator \verb+SIBYLL+ \cite{Engel:2015dxa}.
In this paper we focus on a new evaluation of the prompt flux using next-to-leading
order (NLO) QCD.
We use recently updated PDF's in our pQCD calculations while
limiting the uncertainties in the charm production computation by incorporating
constraints from related high energy collider results.
Finally, we evaluate the prompt neutrino flux and, therefore, the prompt atmospheric
neutrino background for IC by using recent estimates of the cosmic-ray
spectrum and composition \cite{Gaisser:2012zz,Gaisser:2013bla,Stanev:2014mla} and our updated charm production cross-section.

Our focus here is on the link between the charm cross section measurements, including LHC results, and the prompt neutrino
flux in the context of pQCD. This allows us to make estimates of some of the uncertainties in the flux evaluation.
This work is part of a broader program to assess the uncertainties in the theoretical evaluation of the prompt flux
accounting for a wider range of approaches to calculating the charmed hadron cross sections \cite{inprogress}.

In the next section, we review the pQCD evaluation of the production of charmed hadrons.
In section~\ref{sec:prflux}, we describe the $Z$-moment
approach to evaluation the flux, and we show our results for the prompt muon (and electron) neutrino plus antineutrino fluxes.
We briefly discuss the implications of this new flux for the background evaluation at IceCube
in section~\ref{sec:icbkg},
before drawing our conclusions in the final section.

\section{\label{sec:charm-prod}Charm production cross section}

The PeV energy range for atmospheric neutrinos corresponds to an incident energy $E_p\sim 30$ PeV for $pA$ fixed target interactions.
The LHC center of mass energy $\sqrt{s}=7$ TeV is equivalent to a fixed
target beam energy in $pp$ collisions of $E_b=26$ PeV. The LHC measurements of the charm production cross section
\cite{Abelev:2012vra,ATLAScharmconf,Aaij:2013mga} together with
recent RHIC \cite{Adare:2006hc,Adamczyk:2012af}  and modern parton distribution functions (PDFs)
have narrowed down some of the uncertainty in the rate of charm
production in the atmosphere.
The experimental results at high energy for the charm production cross-section in hadronic
collisions are listed in Table~\ref{tab:charmexpts}.

\begin{table}

	\begin{center}
		\begin{tabular}{ l c | c }
			\hline
			\hline
			\noalign{\smallskip}
			Expt.      & ~~~~$ \sqrt{s} $ [TeV]~~~~ & $ \sigma $ [mb]\\
			\noalign{\smallskip}
			\hline
			\noalign{\smallskip}
			PHENIX \cite{Adare:2006hc}
			           & 0.20               & $ \mathbf{0.551}^{+0.203}_{-0.231}\text{ (sys)} $
			\\[10pt]

			%\hline

			STAR \cite{Adamczyk:2012af}
			           & 0.20               & $ \mathbf{0.797} \pm 0.210$
			                                  $ \text{(stat)}^{+0.208}_{-0.295} $
			                                  $ \text{(sys)} $ \\[10pt]
% 			\hline

			\multirow{2}{*}{ALICE \cite{Abelev:2012vra}}
			           & \multirow{2}{*}{2.76}
			                                & $ \mathbf{4.8}\pm 0.8$
			                                  $ \text{(stat)}^{+1.0}_{-1.3}$
			                                  $ \text{(sys)} \pm 0.06\ \text{(BR)}$ \\
			           &                    & $ \pm 0.1 (\text{frag}) \pm 0.1$
			                                  $ \text{(lum)}^{+2.6}_{-0.4}$
			                                  $ \text{(extrap)} $\\[10pt]
% 			\hline

			\multirow{2}{*}{ALICE \cite{Abelev:2012vra}}
			           & \multirow{2}{*}{7.00}
			                                & $ \mathbf{8.5}\pm 0.5$
			                                  $ (\text{stat})^{+1.0}_{-2.4}$
			                                  $ (\text{sys}) \pm 0.1\ (\text{BR})$ \\
			           &                    & $ \pm 0.2 (\text{frag}) \pm 0.3$
			                                  $ (\text{lum})^{+5.0}_{-0.4}$
			                                  $ (\text{extrap}) $\\[10pt]
% 			\hline

			\multirow{2}{*}{ATLAS \cite{ATLAScharmconf}}
			           & \multirow{2}{*}{7.00}
			                                & $ \mathbf{7.13}\pm 0.28$
			                                  $ ({\rm stat})^{+0.90}_{-0.66}$
			                                  $ ({\rm sys}) $ \\
			           &                    & $  \pm 0.78 $
			                                  $ ({\rm lum})^{+3.82}_{-1.90}$
			                                  $ ({\rm extrap}) $\\[10pt]

			LHCb \cite{Aaij:2013mga}
			           & 7.00               & $ \mathbf{6.100} \pm 0.930 $ \\[2pt]
			\hline
			\hline
		\end{tabular}
	\end{center}
	\caption{Total cross-section for $pp(pN)\to c\bar{c}X$ in hadronic
collisions, extrapolated based on NLO QCD by the experimental collaborations from  charmed hadron production measurements in a limited
phase space region.
}
	\label{tab:charmexpts}
\end{table}

In Ref.~\cite{Nelson:2012bc}, Nelson, Vogt and Frawley have investigated a range
of factorization and renormalization scales using the \verb+CT10+ PDF's
\cite{Lai:2010vv} and the NLO order QCD calculation of Nason, Dawson and Ellis \cite{Nason:1987xz,
Nason:1989zy}.
Using a charm quark mass central value of $m_c=1.27$ GeV based on lattice QCD
determinations of the charm quark mass, as summarized in Ref. \cite{Agashe:2014kda},
and a combination of
fixed target, PHENIX, and STAR
charm production cross-sections, they find
that $ M_F/m_c=1.3 $--$ 4.3 $ and $\mu_R/m_c=1.7$--$1.5$  with $M_F=2.1 m_c$ and
$\mu_R=1.6m_c$ as central values. We use these values of parameters
as a guide to
the range of theoretical
NLO charm cross sections expected at high energies.

\begin{figure}[tb]
  \centering
  \includegraphics[width=0.99\textwidth]{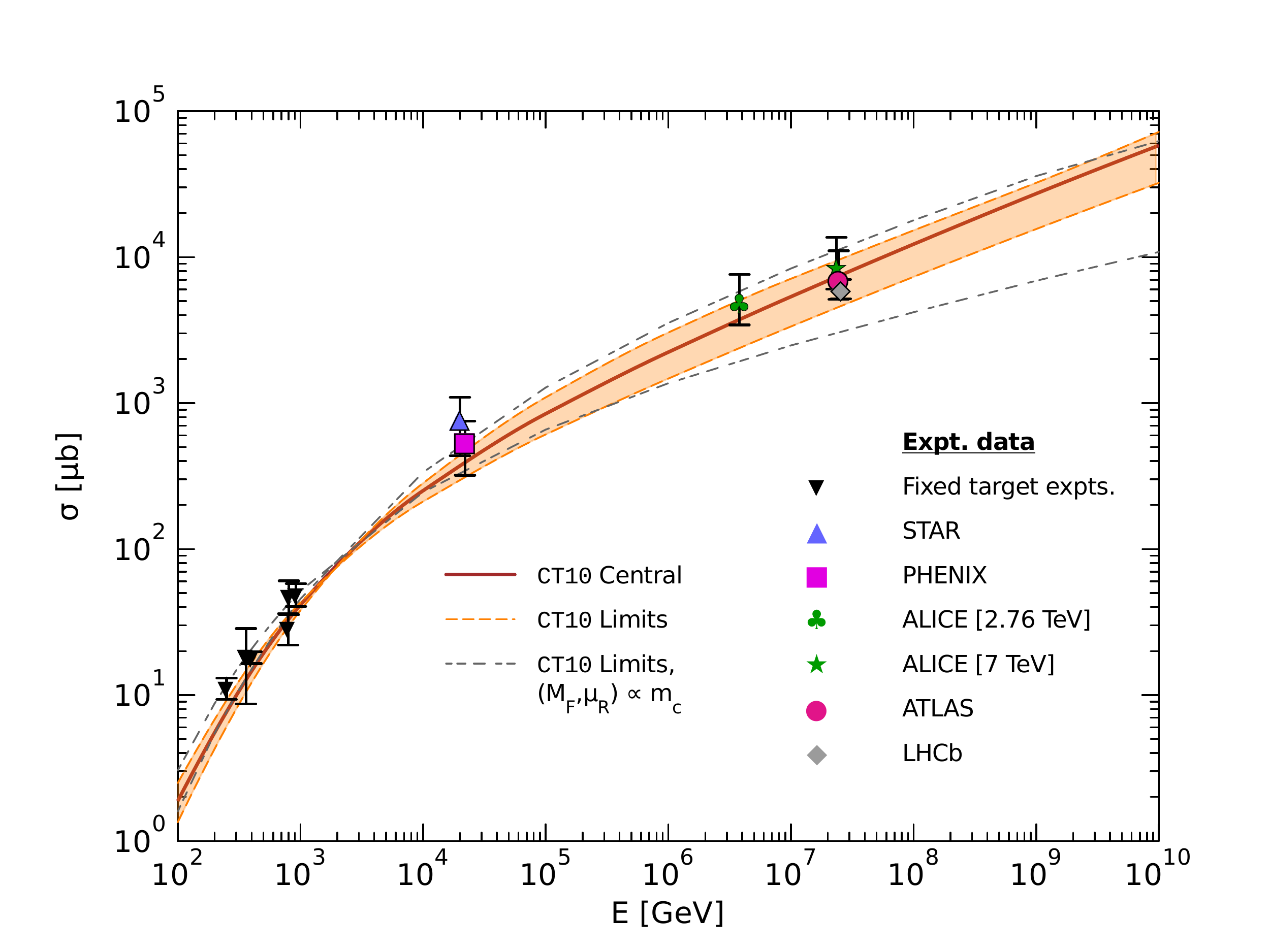}
  \caption{The charm production cross section
           $\sigma_{pN  \to c\bar{c} +X }$ at NLO with $m_c=1.27$ GeV using the
           CT10 parton distributions for a range of scales described in the text,
           with the central set with factorization
           and renormalization scales $M_F=2.10 m_T$ and $\mu_R=1.6 m_T$,
           respectively. Apart from experimental data points listed in
           table~\ref{tab:charmexpts}, results from HERA-B
           \cite{Abt:2007zg}  and lower energy experiments summarized in \cite{Lourenco:2006vw}
           for $ pN $ scattering are shown (labelled as \texttt{Fixed target expts.}).
           For comparison, we also show the lower and upper limits (grey fine-dashed curves) when the
           renormalization and factorization scales are made to vary proportionally to $ m_c $
           rather than to $ m_T $.}
  \label{fig:sigccb}
\end{figure}

In our calculation
we use the NLO Fortran code of Cacciari et al. \cite{Cacciari:1998it,Cacciari:2001td}
that includes the total cross section
\cite{Nason:1987xz}  as well as the single \cite{Nason:1989zy} and double differential
\cite{Mangano:1991jk}
distributions of charm (\ie, $ \diff\sigma/\diff y $ and $ \ddiff \sigma / \diff y\:\diff p_\text{T} $
respectively).
The cross sections shown in figure~\ref{fig:sigccb} for proton interactions
with the iso-scalar nucleon\footnote{Since the charm production cross section is dominated by gluon
fusion,  $p\simeq n\simeq N$. } $ pN \to c\bar{c}X $ are obtained using
factorization and renormalization scales relative to
the charmed quark transverse mass $m_T^2 \equiv (m_c^2+p_T^2)$
rather than relative to the charm quark mass $ m_c$.
This choice of the scale dependence is an acceptable representation of the
factorization and renormalization scales both at moderate energies (where
the choice of scale is not sensitive to $p_T$) and at high energies
(where the $ p_T $ dependence in the scales becomes important).
The \verb+CT10+ PDFs are
our default choice for the NLO evaluations of charm production.
The shaded band in figure~\ref{fig:sigccb}  comes from using $(M_F,\ \mu_R)=(1.25,\ 1.48)\, m_T$
and $(M_F,\ \mu_R)=(4.65,\ 1.71)\, m_T$ (labeled as \verb+CT10+ Limits), as suggested in ref.~\cite{Nelson:2012bc}. The central solid curve (\verb+CT10+
Central) is obtained using $(M_F,\ \mu_R)=(2.1,\ 1.6)\, m_T$. The LHCb \cite{Aaij:2013mga},
ALICE \cite{Abelev:2012vra} and ATLAS \cite{ATLAScharmconf}
cross sections measurements are within our calculated cross section uncertainty band,
thus we use this range of scales for our flux uncertainty band.
The range of QCD parameters consistent with observations have been
studied in \cite{Nelson:2012bc} by taking into consideration
both total as well as differential cross-section measurements at
the LHC for a wide range of the kinematic phase space.
This includes measurements at low transverse momenta and forward rapidities.
The upper and lower limits of our QCD calculation, consistent to within
the error-bars for the experimental data, are based on these parameter
uncertainties (see \eg, shaded area in figure~\ref{fig:sigccb}).

The  total $ pp \to c\bar{c}X $ cross-section at high energies is dominated
by gluon fusion, so the cross section is sensitive to the  gluon PDF small-$x$ behavior.
The perturbative results in \cite{prs} overestimate
the charm cross-section
because of the steeply rising
gluon PDFs at  small-$x$ for
the \verb+CTEQ3+ PDF sets \cite{Lai:1994bb} used, \eg, in \cite{prs,ers}.
By comparison, the updated \verb+CT10+ PDFs we
use here have a more slowly growing
behavior as $ x $ decreases. These PDFs are
fit based on recent experimental inputs
(including from LHC, Tevatron and others).
Using the \verb+MSTW+ \cite{Martin:2009iq} PDFs,
which have an even smaller gluon
PDF growth at small-$ x $, leads to cross-sections that grow even more slowly (compared
to \verb+CT10+) at higher energies, however, the results
still fall within the theoretical uncertainty band shown in figure~\ref{fig:sigccb}.

By choosing $M_F$ and $\mu_R$ with $m_T\to m_c$, we find a similar energy dependence of the total charm cross
section \cite{Nelson:2012bc}.
In addition, the central values for the charm cross sections
are roughly comparable for the same choice of PDF. Using the
\verb+CT10+ PDF and $m_c=1.27$ GeV,
in the energy range $E_b=10^6-10^8$ GeV, the cross section
obtained with $(M_F,\ \mu_R)=(2.1,\ 1.6)\, m_c$ is
within about $\pm 10\%$ of the cross section calculated
with $(M_F,\ \mu_R)=(2.1,\ 1.6)\, m_T$. The
upper choice of scale factors multiplying $m_c$ gives a comparable cross section for the same factors multiplying $m_T$. For the lower range of scale factors, the high energy charm pair production cross section
is smaller for $m_c$ dependent scales.
For example, for $E_b=10^8$ GeV using $(M_F,\ \mu_R)=(1.25,\ 1.48)\, m_c$, the cross section is about
$40\%$ lower than the cross section evaluated using
$m_T$ dependent scales as shown by the grey fine-dashed curves in figure~\ref{fig:sigccb}.
Consequently, while the central values for either choice of the scale give
almost identical prompt neutrino flux, it is the uncertainty band that
is smaller for the
 $\sim  m_T $ scale choice.

As discussed below,  we use the superposition approximation in describing the cosmic ray interactions with air nuclei \cite{Engel:1992vf}. In this approximation, the
cosmic ray all-particle flux is reduced to a nucleon flux.
We require the $NA\to c\bar{c}X$ cross section, and use $ p\simeq n\simeq N $ (as noted previously). 
We continue to label the cosmic ray nucleon with $p$, to distinguish it as the beam nucleon as opposed to the target nucleon.
The average atomic number of the target air nuclei is $A=14.5$. Measurements
of charm production on a variety of targets \cite{Leitch:1994vc,BlancoCovarrubias:2009py}
show that $\sigma(pA\to c\bar{c}X)\simeq A\sigma(pN
\to c\bar{c}X)$. We use this approximation here.

\subsection{Differential cross section}

\begin{figure}[tb]
  \centering
	\includegraphics[width=0.75\textwidth]{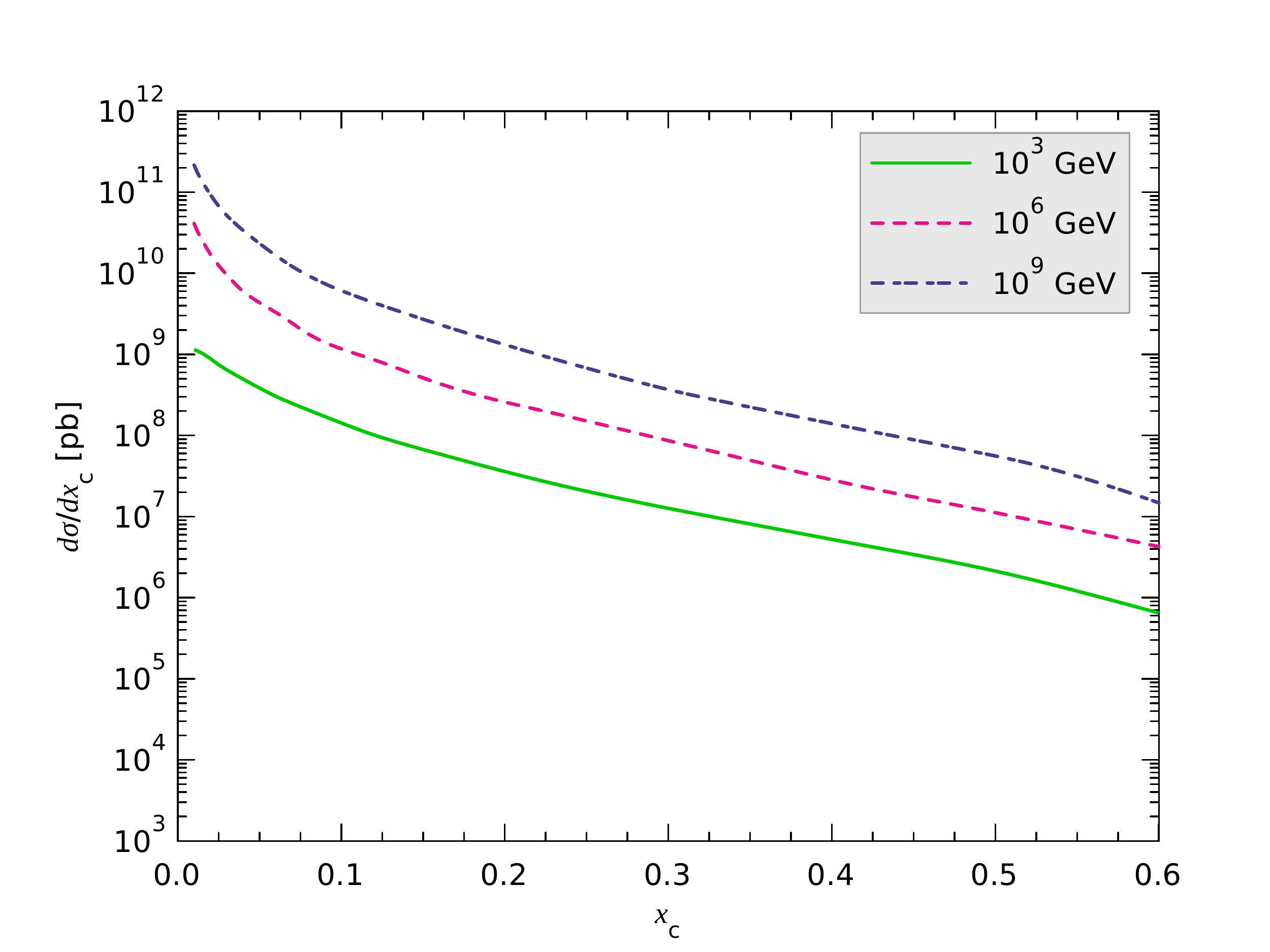}
  \caption{The differential cross section $d\sigma/dx_c$ for the charmed quark, as a function of
           $x_c$ for $E=10^3,\ 10^6, \ 10^9$ GeV for $m_c=1.27$ GeV and
           $M_F=2.1m_T$, $\mu_R=1.6m_T$ using the CT10 NLO PDFs.}
	\label{fig:dsdxe}
\end{figure}

While we seek compatibility of the total charm quark pair production cross section with the results reported by the experimental
collaborations, the dominant contribution to the
prompt flux is from forward production of charm, including fragmentation into charmed hadrons. In our
semi-analytic evaluation of the prompt atmospheric lepton flux, we require the differential
charmed hadron energy distribution,
\begin{equation}
\label{eq:dsdek}
\frac{d\sigma}{dE_h}=\sum_k\int \frac{d\sigma}{dE_k}(AB\to kX)D_k^h\Biggl(\frac{E_h}{E_k}\Biggr)\frac{dE_k}{E_k}
\end{equation}
in terms of the parton level differential distribution and the fragmentation function
$D_k^h$. Here, $h=D^\pm, D^0(\bar{D}^0),D_s^\pm, \Lambda_c^\pm$ and $k=c,\bar{c}$.
We approximate the fragmentation functions for charmed hadrons as energy independent.
Eq. (\ref{eq:dsdek}) can be written as
\begin{equation}
\frac{d\sigma}{dx_E}(pA\to hX)=A\int_{x_E}^{1}
\frac{dz}{z}\frac{d\sigma}{dx_c}(pN\to cX) D_c^h(z)
\end{equation}
in terms of $x_E=E_h/E_b$ and $x_c=E_c/E_b=x_E/z$ for an incident cosmic ray nucleon energy (beam energy) $E_b$.
In figure~\ref{fig:dsdxe}, we show
the differential cross-section as a function of $x_c$ for $E_b=10^3,\ 10^6$ and $10^9$ GeV in $pN$ scattering,
here for $(M_F,\ \mu_R)=(2.1,\ 1.6)\, m_T$. The distributions for $(M_F,\ \mu_R)=(2.1,\ 1.6)\, m_c$
are very similar.

We can compare our results here to those obtained
previously, notably in \cite{prs}. With the \verb+CT10+ NLO PDF's,  the
$m_T$ dependent scales, and a full NLO calculation, we find that our differential distribution at low $ x $
is lower than in ref.~\cite{prs} at high energies (\eg, about 28\% lower at $ 10^9 $ GeV for $ x = 0.1 $).
As previously discussed, this stems from the use of updated PDFs which have
a slower growth at small $x$ than the \verb+CTEQ3+ PDFs used in \cite{prs}.

Our default choice of fragmentation functions for charmed hadrons is that of
Kniehl and Kramer \cite{Kniehl:2006mw}.
The net effect of including the fragmentation functions is
to reduce the predicted flux by about
$\sim 30\%$ relative to the flux without fragmentation included.

\section{\label{sec:prflux}Prompt lepton flux}

We use the $Z$-moment approach \cite{Gaisser:1990vg,Lipari:1993hd}, including an energy dependence of the
$Z$-moments \cite{Gondolo:1995fq} and approximating the depth of the atmosphere as infinite. In the exponential
atmosphere approximation where the density is
\begin{equation}
\rho(h) = \rho_0\exp(-h/h_0)
\end{equation}
for $\rho_0= 2.03\times 10^{-3} $ g/cm$^3$ and $h_0=6.4$ km, the low energy and high energy lepton fluxes have particularly simple forms, involving the spectrum weighted $Z$-moments. The production moments are defined by
\begin{equation}
Z_{ph}(E_h) = \int_{x_{E_\text{min}}}^1\frac{dx_E}{x_E}
              \frac{\phi_p^0(E_h/x_E)}{\phi_p^0(E_h)}\frac{1}{\sigma_{pA}(E_h)}
              \times
              A\frac{d\sigma}{dx_E}(pN\to hX)\ .
\end{equation}
The all-nucleon cosmic ray flux as a function of atmospheric column depth $X$ is
\begin{equation}
\phi_p(E,X)\simeq\phi_p^0(E)\exp(-X/\Lambda_p)=(dN/dE)\exp(-X/\Lambda_p)\ ,
\end{equation}
with $\Lambda_p=\lambda_p(E)/(1-Z_{pp}(E))$. For the proton-air cross section, we use an approximate
parametrization of the EPOS 1.99 cross section \cite{Pierog:2009zt} that is consistent with the high energy
results of the Pierre Auger Observatory \cite{Collaboration:2012wt}.

For decays, the differential cross section is replaced by the differential decay distribution, and the cosmic
ray flux is replaced by the charmed hadron flux.
At high energies, we evaluate the high energy decay $Z$-moment with
a spectral weight of $\phi_p^0(E/x_E)/\phi_p^0(E)$ since $\phi_{h}\propto \phi_p$ at high energies.
The low energy decay $Z$-moment is evaluated using a spectral weight of $1/x_E\cdot\phi_p^0(E/x_E)/\phi_p^0(E)$
since the low energy charmed hadron flux is proportional to $E\phi_p^0(E)$.

For low energy,
the lepton $\ell=\mu,\nu_i$ flux from $h\to \ell$ decays is approximated by
\begin{equation}
\phi_\ell^{low}(h) = Z_{h\ell}^{low}\frac{Z_{ph}}{1-Z_{pp}}\phi_p^0\ ,
\end{equation}
while for high energy (see \eg\ \cite{Gondolo:1995fq}),
\begin{equation}
\phi_\ell^{high}(h)=Z_{h\ell}^{high}\frac{Z_{ph}}{1-Z_{pp}}\frac{\ln(\Lambda_{h}/\Lambda_p)}{1-\Lambda_p/\Lambda_{h}} \phi_p^0\ .
\end{equation}
Each $Z$ factor, $\Lambda$ and $\phi_p^0$ has an energy dependence, suppressed in the notation above.
The resulting lepton flux from charmed hadrons $h$ is
\begin{equation}
\phi_\ell=\sum_{h}\frac{\phi_\ell^{low}(h)\phi_\ell^{high}(h)}{\phi_\ell^{low}(h)+\phi_\ell^{high}(h)}\ .
\label{eq:phinuinterp}
\end{equation}

The lepton fluxes from atmospheric charm depend on the cosmic ray flux directly through $\phi_p^0(E)$ and
in the evaluation of the energy dependent $Z$-moments.

\subsection{Cosmic ray flux}

\begin{figure}[tb]
\centering
  \includegraphics[width=0.9\textwidth]{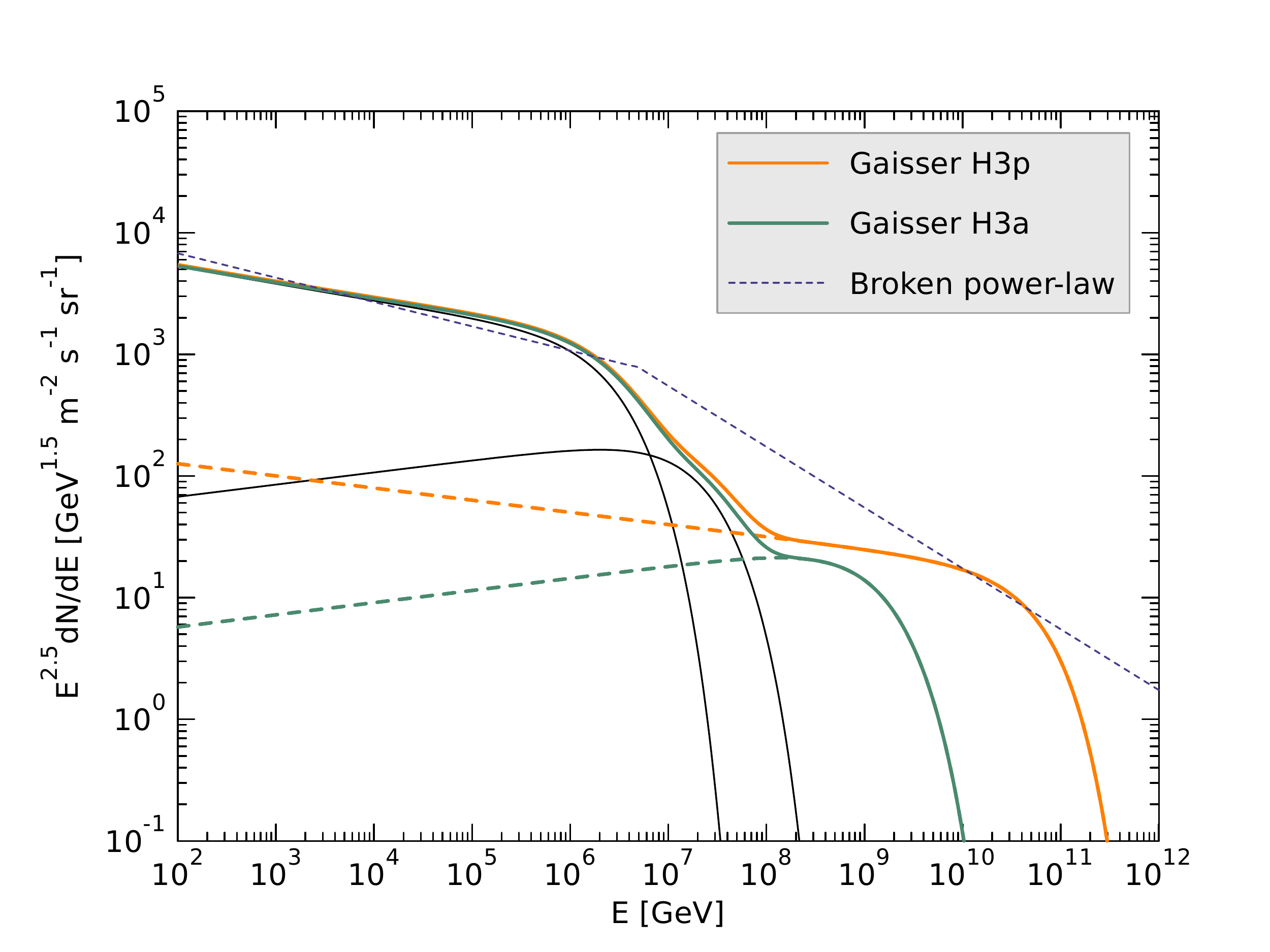}
  \caption{The all-nucleon cosmic ray spectrum as a function of energy per nucleon for the three component
model of ref.~\cite{Gaisser:2012zz} with a mixed extragalactic population (H3a) and all proton extragalactic
population (H3p), and for the broken power-law of eq. (\ref{eq:brokenpl}).}
  \label{fig:crflux}
\end{figure}

The cosmic ray flux has been measured directly and indirectly over a wide
energy range.
Direct measurements are available to energies of about $\sim 100$ TeV.
At higher energies, indirect measurements are made by air shower array
experiments.
While there are some discrepancies in the normalization of the
cosmic ray spectrum at high energies, overall the all-particle cosmic ray
spectrum for the energy range of interest, $10^3$--$10^{10}$ GeV, approximately
follows a broken power-law with the break occurring at $E\simeq 5\times 10^6$
GeV.
Many earlier evaluations of the prompt lepton flux \cite{Gondolo:1995fq,prs,ers,Pasquali:1998xf,Martin:2003us}
used the broken power-law form for the nucleon flux with
\cite{Gondolo:1995fq}:
\begin{align}
\label{eq:brokenpl}
\phi_p^0(E)
=
\begin{cases}
1.7 \, E^{-2.7}  \quad  &\text{for } E<5 \cdot 10^6 \,\, \gev \\
174 \, E^{-3}           &\text{for } E>5 \cdot 10^6 \,\, \gev,
\end{cases}
\end{align}
for $E$  in \gev\ and the nucleon flux in units of
$\text{cm}^{-2} \, \text{s}^{-1} \, \text{sr}^{-1} \, \gev^{-1} $ .
With the fairly recent measurements from ATIC \cite{Panov:2011ak}, CREAM \cite{Ahn:2009tb,Ahn:2010gv}
and Pamela \cite{Adriani:2011cu}, combined with earlier measurements,
Gaisser \cite{Gaisser:2012zz}  and collaborators \cite{Gaisser:2013bla,Stanev:2014mla} have taken
a multicomponent model with three or four source populations to develop models for the cosmic ray
composition. Their parametrizations depend on the particles'
electric charges $Z$ and maximum energies of the source populations, with spectral indices $\gamma$
that vary by population and nucleus.
We use here the parametrization by Gaisser in ref.~\cite{Gaisser:2012zz} with three populations:
from supernova remnants, from other galactic sources and from extragalactic sources.
The H3a flux from ref.~\cite{Gaisser:2012zz} has a mixed composition in
the extragalactic population, while the extragalactic population in what we call the
\emph{H3p flux} is all protons.
Thus, the cosmic ray nucleon spectrum is identical for
H3a and H3p for nucleon energies below $\sim 10^7$ GeV.
Converting  the all-particle flux to the
nucleon flux, the H3a and H3p fluxes are shown along with the broken power-law in figure~\ref{fig:crflux}.

The composition of the cosmic rays causes a much steeper drop in the 
nucleon flux above the knee energy than when
using the simple broken power-law parametrization.
This is particularly important for the high energy prompt lepton flux.
To allow for comparisons with earlier work, we show our results for the prompt lepton flux for the broken power-law
and the H3a and H3p cosmic ray fluxes.

\subsection{$Z$-moment and prompt lepton flux results}

\begin{figure}[tbh]
  \begin{subfigure}{0.49\textwidth}
    \centering
    \includegraphics[width=1.09\textwidth]{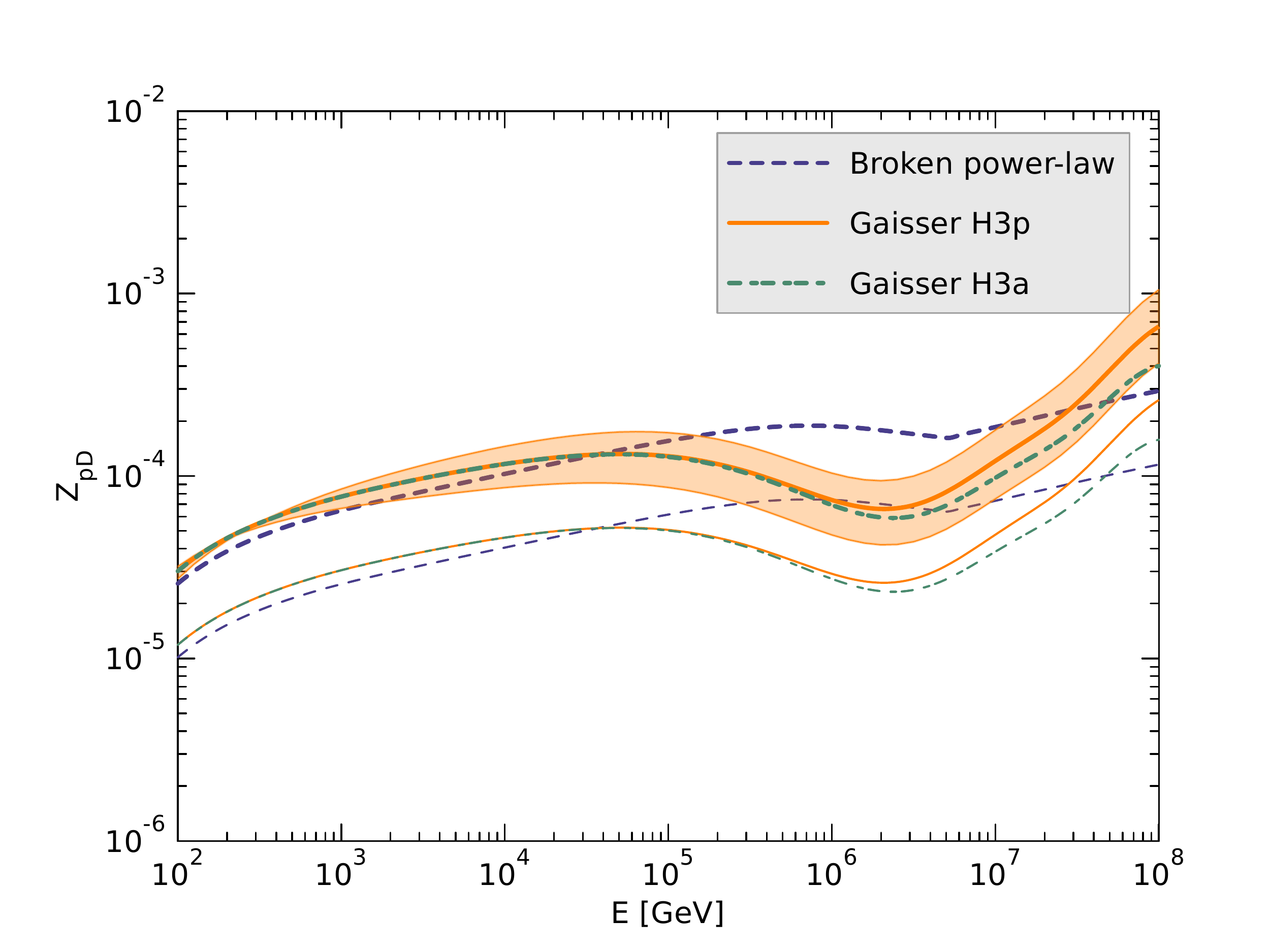}
    \caption{}
    \label{fig:zpd0log}
  \end{subfigure}
  \begin{subfigure}{0.49\textwidth}
    \centering
    \includegraphics[width=1.09\textwidth]{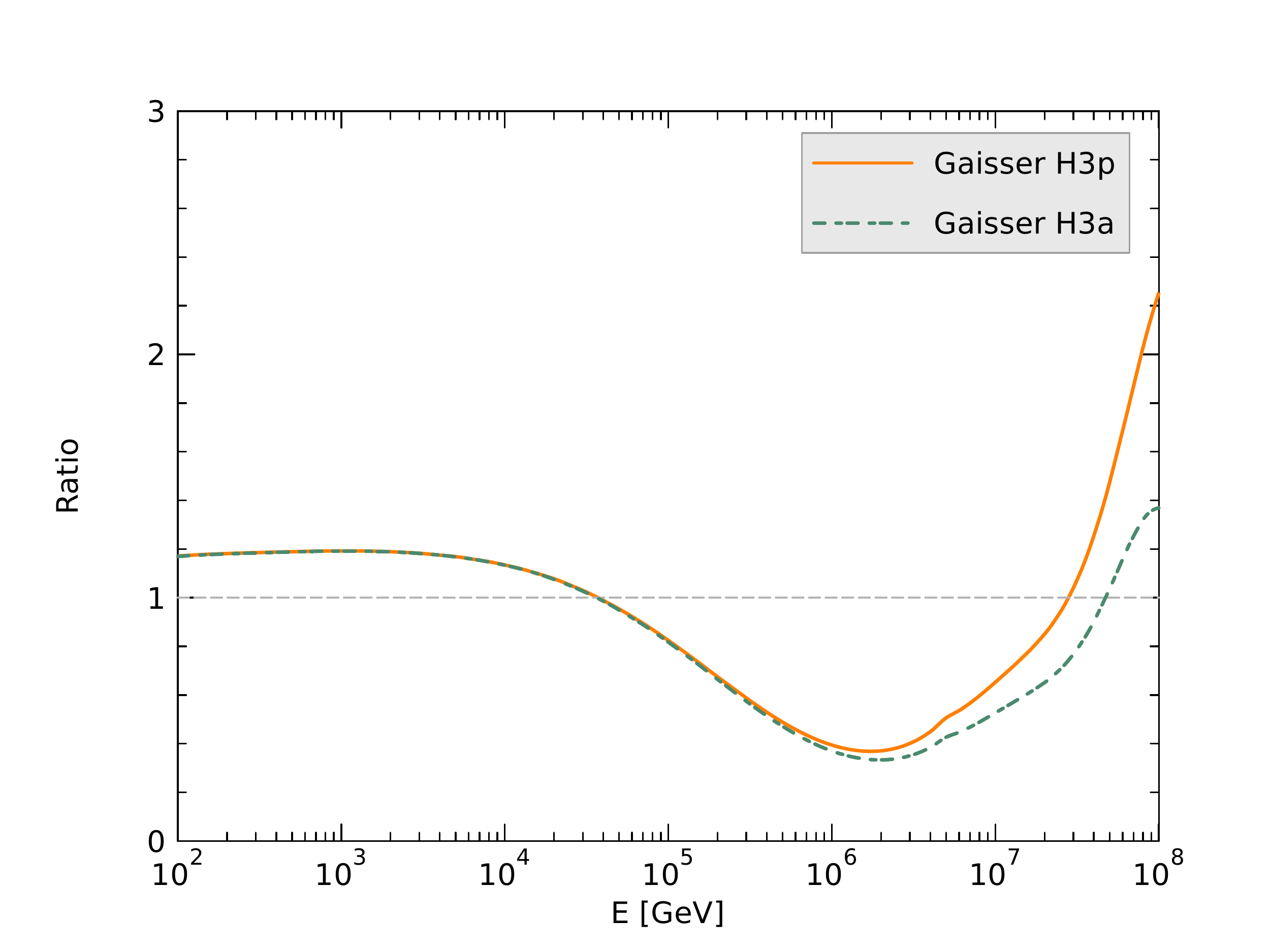}
    \caption{}
    \label{fig:zratio}
  \end{subfigure}
  \caption{(a) Production $Z$-moments for $pA \to M$ for $ M=D^0+\bar{D}^0 $ (as thick curves)
               and $ M = D^{\pm} $ (thin curves) for H3a (green dot-dashed curves), H3p
               (orange solid curves) and broken power-law (blue short-dashed curves) cosmic ray fluxes.
               The range of variation for the \zmom\ is shown (orange shaded region) for the $ Z_{pD^0} $
               when the H3p cosmic ray flux is used for computation, and the relative range of variation
               is identical for the other curves.\\
           (b) Ratio of central \zmom\ for $pA \to D^0+\bar{D}^0$ using the Gaisser H3a and H3p fluxes to the
               broken power-law cosmic ray flux (from eq.~\eqref{eq:brokenpl}).}
  \label{fig:zmoms}
\end{figure}

The production \zmom\ are shown as a function of energy in figure~\ref{fig:zpd0log}.
%This is the sum of both the $D^0$ and $\bar{D}^0$ production.
For each of $ Z_{pD^0} $ and $ Z_{pD^{\pm}} $, the three curves show the moments
evaluated for the three respective cosmic-ray nucleon fluxes presented in figure \ref{fig:crflux}.
For the H3p flux, we also show the band of \zmom\ from the
range of differential cross sections by taking $(M_F,\ \mu_R)=(1.25,\ 1.48)\, m_T$ (for lower limit)
and $(M_F,\ \mu_R)=(4.65,\ 1.71)\, m_T$ (for upper limit).
This relative band of variation is identical for the other \zmom\ shown in the figure.
Figure \ref{fig:zratio} shows the ratio of the central $Z_{pD^0}$-moments obtained using
the H3a and H3p fluxes to that evaluated using the broken power-law nucleon flux.

\begin{figure}[tbh]
  \begin{subfigure}{0.49\textwidth}
    \centering
    \includegraphics[width=1.09\textwidth]{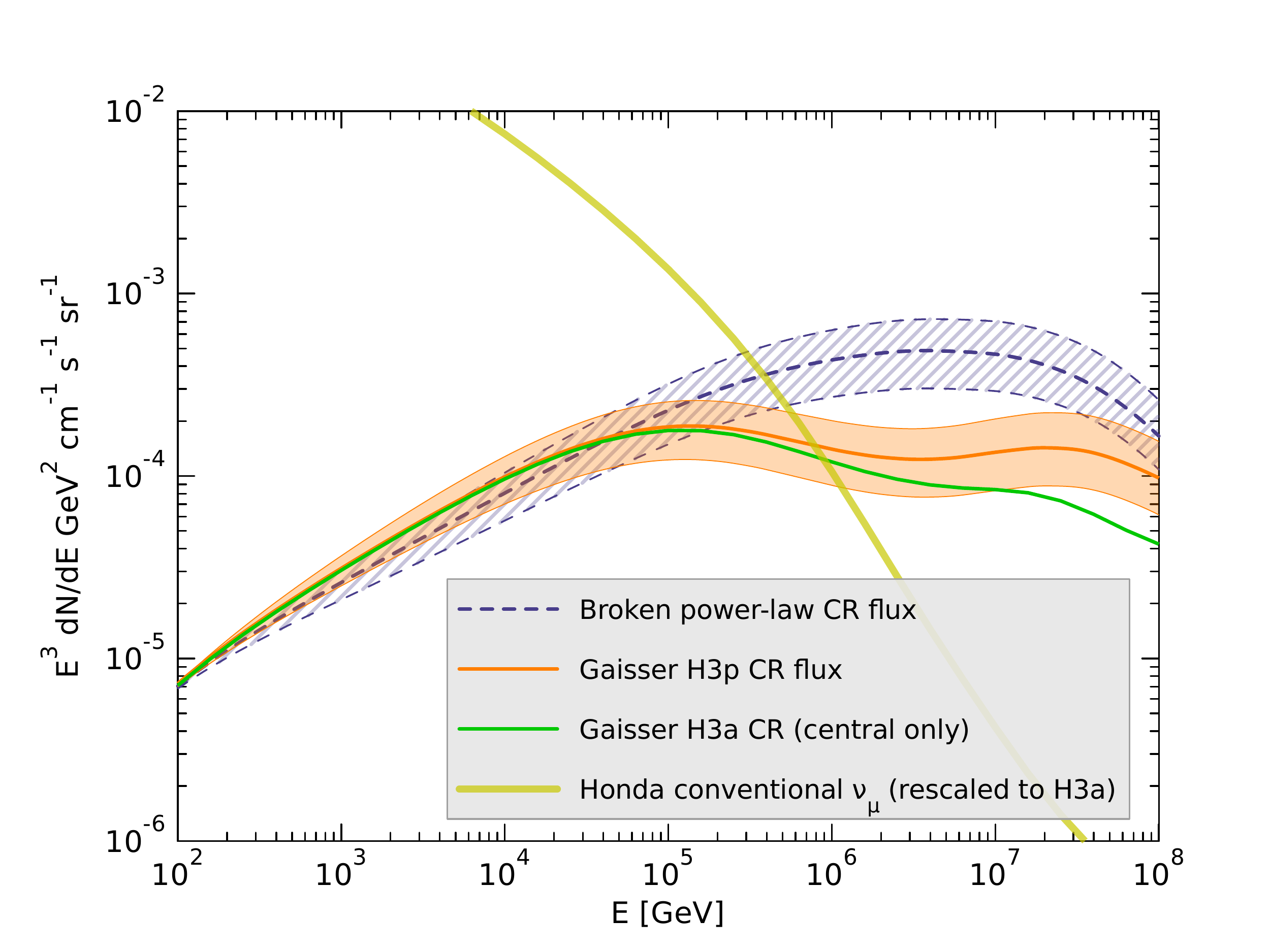}
    \caption{}
    \label{fig:promptiniceflx}
  \end{subfigure}
  \begin{subfigure}{0.49\textwidth}
    \centering
    \includegraphics[width=1.09\textwidth]{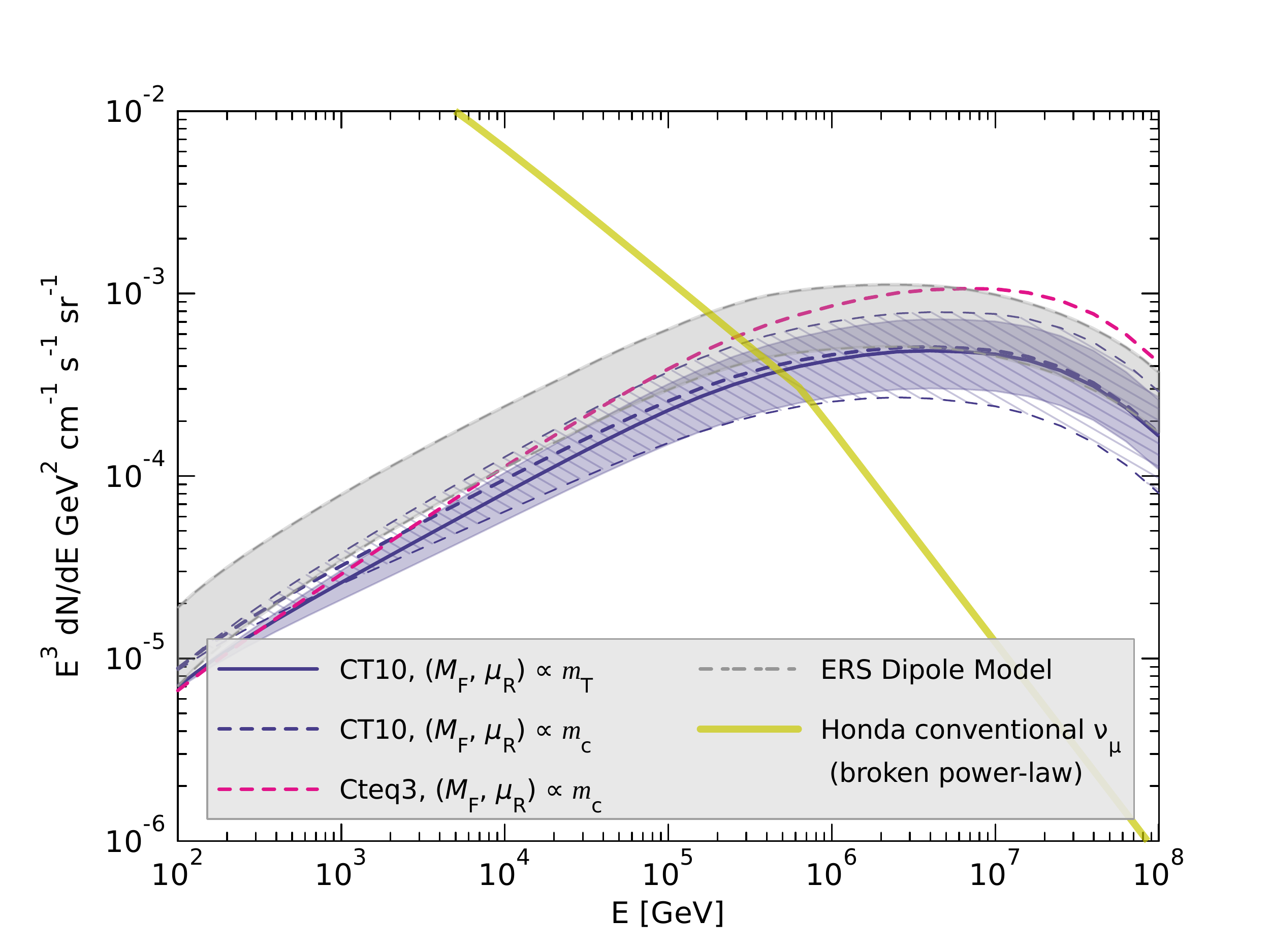}
    \caption{}
	\label{fig:bpl-cmp}
  \end{subfigure}
  \caption{(a) Our benchmark results for the prompt $ \nu_\mu + \bar{\nu}_\mu $
               flux scaled by $ E^3 $ is
               shown as an orange curve, with the cosmic ray flux
               given by the Gaisser H3p fit (see figure \ref{fig:crflux}).
               The blue curve uses instead
               a broken power-law (as used in previous analyses, \eg, \cite{ers}).
               For each curve, the associated shaded region indicates the uncertainty due
               to variation of the QCD parameters.
               The vertical conventional flux from Honda (see, \eg, \cite{Gondolo:1995fq}),
               reweighted to the H3a cosmic-ray primary flux, is also shown.\\
           (b) Comparison of neutrino fluxes with variation in scales and
               PDFs for the broken power-law CR primary flux.
               Shown are the results for central values
               obtained using CT10 as the PDF with
               $ (M_\mathrm{F}, \mu_\mathrm{R}) \propto m_\mathrm{T} $ (solid slate blue line)
               and with scales $ (M_\mathrm{F}, \mu_\mathrm{R}) \propto m_\mathrm{c} $
               (dashed slate blue curve) along with their
               associated bands of variation (corresponding to QCD parameters
               discussed in text) as solid and hatched fills for scales proportional
               to $ m_\mathrm{T} $ and $ m_\mathrm{c} $ respectively.
               The central flux (corresponding to $ M_\mathrm{F} = 2.1m_\mathrm{c}\,,
               \mu_\mathrm{R} = 1.6m_\mathrm{T} $) evaluated for \texttt{CTEQ3} as the PDF is shown
               as the pink dashed curve,
               along with the dipole model computation
               (gray short-dashed curve) of ref.~\cite{ers}.
               The flux uncertainty from \cite{ers} is shown as a grey band.
               For comparison, the vertical conventional flux from Honda (see, \eg, \cite{Gondolo:1995fq}),
               based on the broken power-law cosmic-ray primary flux, is also shown.\\
          }
  \label{fig:promptinice}
\end{figure}

The major difference between the $ D $-meson production \zmom\ when using
the power-law CR flux [from eq.~\eqref{eq:brokenpl}]
against a more recent CR flux estimate, such as the Gaisser H3p flux,
arises at the high energies $ \geqslant 10^5 $ GeV, where the latter
noticeably dip,
before rising sharply at energies beyond the tens of PeV.
In contrast, the \zmom\ when using the broken power-law follows a more steady behavior.
This difference in nature can be traced to the particular
behavior of the Gaisser
cosmic ray primary fluxes---a significant softening of the spectral shape occurs at
around a few PeV energies, where the population transitions from being dominantly galactic to
extra-galactic, before the spectra hardens again at energies around a few hundred PeV
(see figure \ref{fig:crflux}).
When translated to the production \zmom, these effects are visible at comparatively lower energies
because of the inelasticity of the high energy $ pp $ collision, which implies that only a small
fraction (given by $ \langle x_E \rangle \approx 0.1 $) of the incident proton energy goes into the produced $ c\bar{c} $.
The nature of the \zmom, in turn, translates directly to the total prompt lepton flux (as shown
in figure \ref{fig:promptiniceflx}).
The central \zmom\ obtained using the H3p estimate will henceforth be our benchmark result
when determining the prompt flux and correspondingly the event-rates at IC.

As discussed above, we use the charmed hadron spectral weights for the decay \zmom. These are evaluated using
$dn/dE$ from ref.~\cite{Gaisser:1990vg,Lipari:1993hd,Bugaev:1998bi}.

Additional $Z$-moments are needed for the flux evaluation, in particular $Z_{pp}$ and $Z_{hh}$ along with
$\lambda_h$. For $Z_{pp}$, we have approximated the $pA\to pX$ differential cross section with a scaling
form
\begin{equation}
\label{eq:pA-dsigma}
\frac{d\sigma}{dx_E}\simeq \sigma_{pA}(E)(1+n)(1-x_E)^n
\end{equation}
with $\sigma_{pA}$ as described above and $n=0.51$. With these choices, at $E=10^3$ GeV for the broken
power-law, $Z_{pp}=0.271$ and $\Lambda_p=\lambda_p/(1-Z_{pp})=116$ g/cm$^2$. By comparison,
the scaling values in \cite{Gaisser:1990vg,Lipari:1993hd} are $Z_{pp}=0.263$ and $\Lambda_p=117$ g/cm$^2$. The change in the cosmic ray spectrum with the broken power-law together with the energy
dependence of the cross section reduces our calculated
$Z_{pp}$ to 0.231 and $\Lambda_p$ to 67 g/cm$^2$ at $E=10^8$ GeV. Similar results are obtained for the H3a
and H3p cosmic ray flux inputs.
We remark that in \cite{Gondolo:1995fq}, energy dependent \zmom\ evaluated using the \verb+PYTHIA+ Monte Carlo program \cite{Sjostrand:1993yb} are used, giving \eg\ $Z_{pp}(10^3\ {\rm GeV})\simeq 0.5$. These are also used in \cite{ers}.
The low energy flux is proportional to $(1-Z_{pp})^{-1}$, so this numerical factor is important to the overall normalization.

For the charmed hadrons' interaction lengths and interaction $Z$-moments, we use kaon-proton interactions as representative. For all charmed hadrons, we use the same expressions, based on kaons.
We take
\begin{equation}\label{eq:dsigma-air}
\frac{d\sigma}{dx_E}\simeq A^{0.75}\sigma_{KN}(E)(1+n)(1-x_E)^n
\end{equation}
with $\sigma_{KN}$ determined by the COMPAS group and
summarized by the Particle Data Group in \cite{Agashe:2014kda}.
We find that
setting $ n=1 $ gives $ Z_{KK}=0.217 $ and $\Lambda_K=162$ g/cm$^2$ for the broken power-law at $10^3$ GeV,
reducing to $Z_{KK}=0.176$ and $\Lambda_K=40$ g/cm$^2$ at $10^8$ GeV.
The scaling values in \cite{Gaisser:1990vg,Lipari:1993hd} are 0.211 and 175 g/cm$^2$, respectively.
The precise value of $n$ for meson scattering in eq.~\eqref{eq:dsigma-air} affects only  $\phi_\ell^{high}$.

Our prompt lepton fluxes are shown in figure~\ref{fig:promptiniceflx}. We show
$E^3\phi_{\nu_\mu+\bar{\nu}_\mu}$ as a function of neutrino energy. The fluxes of
$\mu+\bar{\mu}$ and $\nu_e+\bar{\nu}_e$ are the same as shown in the figure, in the approximations
we use here. The upper band shows our NLO
result using \verb+CT10+ PDFs with the range of $(M_F,\mu_R)$ discussed in section~\ref{sec:charm-prod},
using the broken power-law as the input cosmic ray all-nucleon spectrum.
The lower band shows the prompt flux
using the H3p cosmic ray flux inputs.
The H3a cosmic ray flux results in a lower prompt lepton flux for
energies above $\sim 2\times 10^5$ GeV, roughly a factor of two lower at $E=10^8$ GeV.

\subsection{Comparison with previous prompt neutrino flux calculations}

In \cite{ers}, the dipole model approach including effects of parton saturation 
was used for charm production in
$ pp \to c\bar{c}X $.
With respect to the total neutrino flux obtained therein, we
find that our benchmark results, \ie, obtained using Gaisser
H3p cosmic ray flux
as opposed to the use of broken power-law spectra in the former,
 yields prompt neutrino fluxes
that are reduced by a factor which varies from about 2 at
 lower energies (below 100 TeV) to a maximum reduction factor of about 8 at high energies (few PeV).
On the contrary, the result for the
central value of the flux obtained here using a
broken power-law cosmic ray spectra, for
comparison, is seen to be
lower compared to the results in \cite{ers} by up to a factor of 3.
This reduction, despite the use of the same cosmic ray spectra,
mainly arises due to the updated evaluation of the
$ Z_{pp} $ moment [as explained just below eq.~\eqref{eq:pA-dsigma}], and from 
differences  in the large $x_E$ values of $d\sigma/dx_E$ in the 
dipole and perturbative QCD approaches.
Figure~\ref{fig:bpl-cmp} additionally shows the
flux that arises for the central scale choice performed
with the \verb+CTEQ3+ PDFs and the broken power-law in
eq.~\eqref{eq:brokenpl}.
We find that the steep increase in cross-section
with energy that one obtains when using
this older PDF set, translates directly to larger fluxes at energies of 10 TeV and higher
(compared to central \verb+CT10+ fluxes obtained using the same cosmic-ray primary flux);
nonetheless, it turns out to be smaller
at low energies than that obtained in the
dipole computation in \cite{ers}.
Finally, figure~\ref{fig:bpl-cmp} also shows that using scales proportional to
$ m_\mathrm{c} $ rather than
$ m_\mathrm{T} $ leads to similar central predictions but
somewhat larger bands of uncertainty.
This is a feature that can be traced back to the $ pp \to c\bar{c}X $ cross-section,
as discussed in section~\ref{sec:charm-prod}.

For both figures \ref{fig:promptiniceflx} and \ref{fig:bpl-cmp}, we show the conventional
atmospheric neutrino flux from ref.~\cite{Honda:2006qj}.
Since the original computation of this flux in \cite{Honda:2006qj} was based on the broken
power-law cosmic ray primary flux, for consistency, we reweigh this flux in
figure~\ref{fig:promptiniceflx} to account for the updated H3a cosmic-ray primary flux\footnote{The
difference between results obtained using H3a and those using H3p fluxes is negligible below 1 PeV.
Thus, we do not distinguish between the two when computing the conventional flux, which is only
significant at much lower energies, \ie, up to 200 TeV.}
using the code \verb+NeutrinoFlux+ \cite{neutrinoflux}.
This reweighted flux is used also to determine the resulting conventional atmospheric
background in figure~\ref{fig:ICevents}.

Quantifying, approximately, the changes in our evaluation over that in \cite{ers} aside from
using perturbative QCD rather than the dipole model approach,
we note that the following factors are responsible for a reduced prompt flux:
\begin{itemize}
  \item Use of recent cosmic-ray flux estimates which predict a lower nucleon flux at  high energies
        ($ 10^7 $--$ 10^9 $ GeV) than the previous broken power-law (see eq.~\eqref{eq:brokenpl}).
        Correspondingly, the prompt flux evaluated using the H3p cosmic ray primary flux is reduced
        by a factor of 1.2--3 (figure~\ref{fig:promptiniceflx}) relative to the broken power-law
        between the energies $ 10^5 $--$ 10^6 $ GeV,
        for example, where the prompt background is expected to be important.
  \item Reevaluation of $ Z_{pp}\text{ and } \Lambda_{p} $, and
        $ Z_{kk}\text{ and } \Lambda_{K} $ , using the H3p cosmic ray flux and a parametrization of the $p\to p$,
      $D\to D\simeq K\to K$ differential cross sections, and using %based on improved fits
         an updated
        $ p $-Air cross-section, decreases the flux by about 30\%.
  \item Use of \verb+CT10+ PDFs instead of the older \verb+CTEQ3+ PDFs used in \cite{prs} reduces
        the flux by an additional factor of 2--3 at the high energies
        ($ \gtrsim 10^5 $ GeV), as figure~\ref{fig:bpl-cmp} shows by comparing
        fluxes computed using the same broken power-law cosmic-ray primary flux
        and QCD parameters, but with the two different PDF sets.
  \item Between the dipole calculation in \cite{ers} and the results shown here for 
        the broken power law cosmic ray flux, the large $x$ behavior of the 
        dipole and perturbative QCD evaluations of $pA\to c\bar{c}X$, input to 
        the evaluation of the production $Z$-moments,  account for the relative 
        fluxes differing by a factor of $\sim 1.5$.
        Uncertainties associated with variations of the QCD parameters for the
        perturbative and dipole evaluations are shown with the (overlapping)
        bands in figure~\ref{fig:bpl-cmp}.
        
\end{itemize}

\section{\label{sec:icbkg}Prompt background at IceCube}

\begin{figure}[htb]
  \centering
  \includegraphics[width=0.95\textwidth]{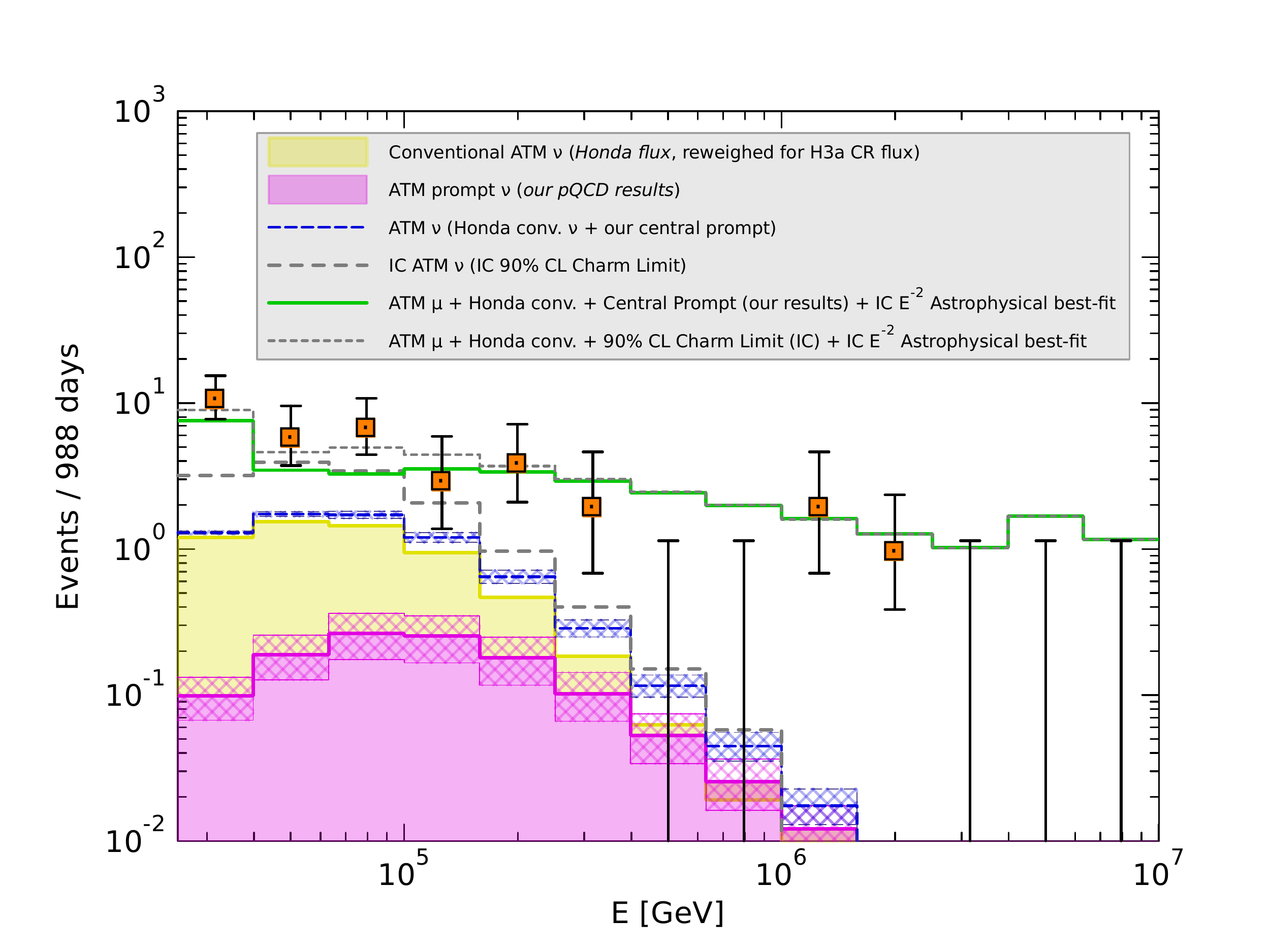}
  \caption{Event rates at IceCube from prompt neutrinos, with our updated prediction
           for the prompt flux indicated in magenta, along with uncertainties from variation
           in the QCD parameters indicated as a hatched region around the central curves.
           The re-evaluated total atmospheric neutrino background (blue dashed curve) includes
           events from our central prompt prediction and the Honda conventional neutrino flux reweighted
           for H3a cosmic ray flux (with the uncertainty in prompt event rates indicated as a hatched
           region around it).
           The total neutrino background estimated by IC at the level of 90\% CL charm limit
           \cite{Aartsen:2014gkd} is shown (dashed gray curve) for comparison.
           The prediction for total event rates using the $ E^{-2} $ fit astrophysical
           signal from \cite{Aartsen:2014gkd} and updated atmospheric background
           (\ie, including our re-evaluated prompt background) is shown as a green curve,
           while the similar IC estimate using the older atmospheric background and prompt at
           the level of 90\% CL (see \cite{Aartsen:2014gkd}) is shown as the gray thick-dashed curve.
           For both these latter curves, the background includes contribution from atmospheric
           muons (reproduced from \cite{Aartsen:2014gkd}), in addition to atmospheric neutrinos.
           Observed total event rates at the IC are shown as solid red blocks, along with their associated
           $ 1\sigma $ statistical uncertainty.}
	\label{fig:ICevents}
\end{figure}

The IceCube
atmospheric neutrino background event rates, both prompt and conventional, can be
evaluated from the corresponding flux by making use of the flavor-dependent
effective areas for the detector, given in \cite{IceCube}.
The central value of the estimated prompt neutrino background relevant to the
988-day IC observation is shown in figure \ref{fig:ICevents}.
Additionally, the figure also shows the maximum and minimum estimates due to
variation in the cross-section (see figure \ref{fig:sigccb})  and differential
charm quark energy distribution obtained by varying
the QCD parameters compatible with
experimental results.\footnote{When comparing
event rates from prompt neutrino flux obtained here to
those obtained in \cite{Aartsen:2014gkd}, the difference is not
as stark as
one would expect from a straight comparison
at the level of fluxes obtained here to that
obtained in \cite{ers}, because IC implements
a reweighing procedure on the central estimate of the
prompt neutrino flux in \cite{ers} to already account
for updated cosmic-ray
spectra.}
We note that the procedure using the IC effective areas overestimates 
the prompt event rates because it does not take into account the 
neutrino self-veto as discussed in \cite{Gaisser:2014bja}.

Using the Honda estimates for conventional atmospheric neutrino flux,\footnote{For
the conventional atmospheric neutrino flux too, IC uses the
Honda flux reweighted to account for changes in fits
to the cosmic-ray spectra from older power-law estimates to more recent estimates, \eg\ in
\cite{Gaisser:2012zz}.
We have used the same method in figure~\ref{fig:promptiniceflx} to obtain the conventional neutrino flux.}
events from
background muons estimated in \cite{Aartsen:2014gkd},
and our central prompt flux, we compute the total atmospheric background.
This is shown in figure \ref{fig:ICevents} as the blue dashed curve,
with the uncertainty indicated as a hatched area around this central curve.
When compared with the conventional atmospheric neutrino background,
the prompt flux is only a minor contributor at energies where the
background is significant (about 200 TeV and lower).
Consequently, the total atmospheric
neutrino background at these energies is not significantly affected by the
$ \sim $ 10\%--15\% uncertainty in the prompt flux from variation of the QCD parameters.

It is clear from a comparison with the total atmospheric background (gray dashed curve)
in figure \ref{fig:ICevents}, obtained using a previous results for the prompt neutrino flux
\cite{Aartsen:2014gkd} and IC's experimental 90\% CL prompt flux limits,\footnote{IC's 90\% CL limits
on the prompt flux corresponds to 3.8 times the central prompt flux estimated in \cite{ers}.}
that our new pQCD results
lead to a significant reduction in the total background in the lower energies up to 100 TeV.
This will in turn lead to revisions in the evaluation of statistical
significances of the astrophysical signal for the IC, but a detailed
re-analysis of the IC signal is beyond the scope of the present work.
We note that this reduced background implies that the IC fixed-slope
$ E^{-2} $ best-fit astrophysical flux
underestimates the low-energy signal in the observed events.

\section{\label{sec:conc}Conclusions}

Hadronic interactions at extremely high energies necessarily involves understanding
of QCD at these energies; however, this is beset by relatively large uncertainties
in the related computation.
Interactions between cosmic ray protons and atmospheric nuclei which lead to the
production of neutrinos via the formation of a mesons ($ \pi^{\pm},\ K^{0,\pm},
\ D^{0,\pm} ,\  D_s^\pm$) and baryons, and its consequent decay is one such process.
In the era of the IceCube observations of ultra-high energy neutrinos, the importance of properly
understanding and estimating the background from atmospheric neutrinos cannot be
overstated.
Since earlier perturbative QCD results for prompt neutrino fluxes
were based on fairly old  PDFs and were not constrained by the
recent LHC data, we have revisited the
computation in the present work, incorporating
recent developments in the understanding
of low-$ x $ PDF's and using
recent experimental results that have bearing on the
relevant QCD inputs.

By varying the QCD parameter space including the involved factorization and renormalization
scales within bounds set by results from LHCb, ALICE, ATLAS among others, and making use of
recent \verb+CT10+ PDF's, we have re-evaluated $ \sigma(pp\to\ccb X) $.
We have determined $ \sigma(pN \to \ccb X) $ from the latter, and thereafter, using
the standard procedure of energy-dependent $Z$-moments, we have computed the total
prompt flux expected at the detector.
For an estimate of the composition and spectrum of the cosmic ray primaries, we have relied on
the recent results from ref.~\cite{Gaisser:2012zz} rather than use a broken power-law
which is now known to be an inaccurate representation of the cosmic ray nucleon flux at the
high energies.

We find that the prompt neutrino background at the IceCube is
% much
lower than that estimated in \cite{IceCube}.
We note that this may lead to a consequent revision in the estimates of the
statistical significance of the IC signal over the now reduced atmospheric
background.

\acknowledgments
We thank the IceCube 
collaboration for providing their code \verb+NeutrinoFlux+ that we used to 
compute the reweighted conventional neutrino fluxes, and David Boersma, Olga 
Botner, Teresa Montaruli and Anne Schukraft for help with the code.
This research was supported in part by the US Department of
Energy through contracts DE-FG02-04ER41319, DE-FG02-04ER41298,
DE-FG03-91ER40662, DE-FG02-13ER41976, DE-SC0010114 and DE-SC0002145,
in part by the National Science Foundation
under Grant No.\ NSF PHY11-25915, in part by the
Polish NCN grant DEC-2011/01/B/ST2/03915, and in part by the Swedish Research Council 
under contracts 2007-4071 and 621-2011-5107.
We thank Nordita and KITP for the
hospitality while part of this work was completed.

\bibliographystyle{JHEP}
\bibliography{refs}

\end{document}